\documentstyle[psfig,12pt,axodraw]{article}
\def\lsim{\raise0.3ex\hbox{$\;<$\kern-0.75em\raise-1.1ex\hbox{$\sim\;$}}}
\def\gsim{\raise0.3ex\hbox{$\;>$\kern-0.75em\raise-1.1ex\hbox{$\sim\;$}}}
\def\lsim{\:\raisebox{-0.5ex}{$\stackrel{\textstyle<}{\sim}$}\:}
\def\gsim{\:\raisebox{-0.5ex}{$\stackrel{\textstyle>}{\sim}$}\:}
\def\half{{\textstyle{1 \over 2}}}
\def\ie{{\it i.e.}}

\begin{document}
\begin{titlepage}
\begin{flushright}
hep-ph/9806382\\
FTUV/98-50\\
IFIC/98-51\\
\today
\end{flushright}
\vspace*{5mm}
\begin{center} 
{\Large \bf Tau Lepton Mixing with Charginos and its Effects on
Chargino searches at $e^+e^-$ Colliders}
\\[15mm]
{\large A.G. Akeroyd, Marco A. D\'\i az and Jos\'e W.F.  Valle}
\\
\hspace{3cm}\\
{\small Departamento de F\'\i sica Te\'orica, IFIC-CSIC, Universidad
 de Valencia}\\ 
{\small Burjassot, Valencia 46100, Spain}
\end{center}
\vspace{5mm}
\begin{abstract}
In bilinear R--Parity violating models where a term $\epsilon_3L_3H_2$
is introduced in the superpotential, the tau lepton can mix with
charginos. We show that this mixing is fully compatible with LEP1
precision measurements of the $Z\tau\tau$ and $W\tau\nu_{\tau}$
couplings even for large values of $\epsilon_3$ and of the induced
vacuum expectation value $v_3$ of the tau--sneutrino.  The single
production of charginos at $e^+e^-$ colliders is possible in this case
and we present numerical values of the cross--section at LEP1, LEP2 and
an NLC. We find maximum values of 10 pb at LEP1 and 1 fb at NLC, while
the corresponding values at LEP2 are too small to observe.
\end{abstract}

\end{titlepage}

\setcounter{page}{1}

\section{Introduction}

The Minimal Supersymmetric Standard Model (MSSM) \cite{mssm} is the
most popular extension of the Standard Model (SM) and its
phenomenology has received much attention in recent years. The MSSM
assumes the conservation of R-parity \cite{RP}, a discrete symmetry
given by $R_p=(-1)^{(3B+L+2S)}$, where L is the lepton number, B is
the baryon number and S is the spin of the state. Two important
consequences of such a symmetry are that all supersymmetric particles
must be pair-produced, with the lightest of them being stable. In
recent years growing attention has been given to models in which the
conservation of R-Parity is relaxed \cite{rpv} by adding explicit
R-parity violating terms in the superpotential. In the literature one
finds many studies using models which posess cubic Rp violating
superpotential terms, which in turn introduce a very large number of
arbitrary Yukawa couplings \cite{3lin}.  Such models suffer from the
fact that their phenomenological study involves a large number of free
parameters. Some of these must be strongly suppressed in order to
avoid conflict with nucleon stability. In this work we shall first
focus on Bilinear R--Parity Violation (BRpV) which has been advocated
in a number of previous papers
\cite{e3others,chaHiggsEps,epsrad,yukeps}. Such models allow one to
map out the phenomenological potential of the model in a systematic
fashion \cite{beyonddesert,epstalks}.  They have the theoretical
advantage of being a low energy approximation of spontaneous R--parity
violating models. These may have either gauged \cite{ConchaV} or
ungauged \cite{finnish} lepton number. In the latter case the particle
spectrum will include a massless majoron \cite{MV_RIV}. The presence
of such a particle allows one to avoid the stringent cosmological
nucleosynthesis bound of $m_{\tau_{\nu}}\le 1-2$ MeV or so
\cite{beyonddesert}, and instead employ the weaker mass limit
found from direct searches at LEP1, $m_{\tau_{\nu}}\le 18$ MeV.
However, in order to avoid excessive stellar cooling through majoron
emission the value of the sneutrino vacuum expectation value ($v_3$)
is then constrained to small values ($\le 100$ MeV) \cite{MV_RIV}. In
contrast, in the simplest model which {\sl explicitly} breaks
R--parity via a bilinear superpotential term $v_3$ is only constrained
by the $W$ mass formula, and may be comfortably varied up to values of
order 100 GeV.

The Tau lepton in the BRpV or in spontaneous Rp breaking models is
allowed to mix with the charginos, an effect dependent on the values
of $\epsilon_3$ and $v_3$. Such mixing, if substantial, would naively
be expected to affect the theoretical value of the $Z\tau\tau$
coupling which is measured to very high accuracy at LEP1 (error$\approx
0.25\%)$. Thus it is important to check if one may constrain the
allowed parameter space of $\epsilon_3$ and $v_3$, and we do this in
the first part of this paper. A consequence of the above mixing is the
possibility of single chargino production at $e^+e^-$ colliders, a
process not possible in Rp conserving models. We update and improve
previous analyses \cite{ConchaV,chitau} of this channel at LEP1 and
investigate the phenomenology at LEP2 and a NLC.  The paper is
organized as follows. In Section 2 the essential properties of the
BRpV model are reviewed and we present the relevant mass matrices.  In
Section 3 we study the effect of the chargino--tau mixing on the
$Z\tau\tau$ and $W\tau\nu_{\tau}$ couplings, while Section 4 analyses
the prospects of single chargino production at $e^+e^-$ colliders.
Section 5 contains our conclusions.
 
\section{The Model}

In the simplest BRpV model the supersymmetric Lagrangian is specified
by the superpotential $W$ given by
\begin{equation} 
W=\varepsilon_{ab}\left[
 h_U^{ij}\widehat Q_i^a\widehat U_j\widehat H_2^b
+h_D^{ij}\widehat Q_i^b\widehat D_j\widehat H_1^a
+h_E^{ij}\widehat L_i^b\widehat R_j\widehat H_1^a
-\mu\widehat H_1^a\widehat H_2^b
+\epsilon_i\widehat L_i^a\widehat H_2^b\right]
\label{eq:Wsuppot}
\end{equation}
where $i,j=1,2,3$ are generation indices, $a,b=1,2$ are $SU(2)$
indices. The last term in eq.~(\ref{eq:Wsuppot}) is the only one which
violates $R$--parity. Such a term arises in spontaneous R--parity 
breaking models, with $\epsilon_i$ generated by the product of a new 
Dirac-type neutrino Yukawa coupling and some new vacuum expectation 
value (VEV) of an SU(2) singlet sneutrino field. The parameters 
$\epsilon_i$ have the dimension of mass, and for simplicity we consider 
only $\epsilon_3$ non--zero. Of particular interest to us is the
chargino/tau mass matrix, which is given by
\begin{equation}
{\bf M_C}=\left[\matrix{
M & {\textstyle{1\over{\sqrt{2}}}}gv_2 & 0 \cr
{\textstyle{1\over{\sqrt{2}}}}gv_1 & \mu & 
-{\textstyle{1\over{\sqrt{2}}}}h_{\tau}v_3 \cr
{\textstyle{1\over{\sqrt{2}}}}gv_3 & -\epsilon_3 &
{\textstyle{1\over{\sqrt{2}}}}h_{\tau}v_1}\right]
\label{eq:ChaM6x6}
\end{equation}
and $M$ is the $SU(2)$ gaugino soft mass. The form of this matrix is
common to all models with spontaneous breaking of Rp, as well as in
the simplest truncation of these models provided by the BRpV model.
We note that the chargino sector decouples from the tau sector in the
limit $\epsilon_3=v_3=0$.  As in the MSSM, the chargino mass matrix is
diagonalized by two rotation matrices $\bf U$ and $\bf V$
\begin{equation}
{\bf U}^*{\bf M_C}{\bf V}^{-1}=\left[\matrix{
m_{\chi^{\pm}_1} & 0 & 0 \cr
0 & m_{\chi^{\pm}_2} & 0 \cr
0 & 0 & m_{\tau}}\right]\,.
\label{eq:ChaM3x3}
\end{equation}
The lightest eigenstate of this mass matrix must be the tau lepton
($\tau^{\pm}$) and so the mass is constrained to be $1.777$ GeV. 
To obtain this the tau Yukawa coupling becomes a function of the 
parameters in the mass matrix, and the full expression is given in 
\cite{chaHiggsEps}. The composition of the tau is given by:
\begin{equation}
\tau^+_R=V_{3j}\psi_j^+, \qquad \tau^-_L=U_{3j}\psi_j^- 
\label{taucomp}
\end{equation}
where $\psi^{+T}=(-i\lambda^+,\widetilde H_2^1,\tau_R^{0+})$ 
and $\psi^{-T}=(-i\lambda^-,\widetilde H_1^2,\tau_L^{0-})$. The
two component Weyl spinors $\tau_R^{0-}$ and $\tau_L^{0+}$ are weak 
eigenstates and, similarly, the two component Weyl spinors $\tau^+_R$ 
and $\tau^-_L$ are mass eigenstates.

It follows easily from eq.~(\ref{eq:ChaM3x3}) that the matrix 
$\bf{M_CM_C^T}$ is diagonalized by $\bf U$ and the matrix $\bf{M_C^TM_C}$ 
is diagonalized by $\bf V$. It is instructive to write the matrices 
$\bf{M_CM_C^T}$ and $\bf{M_C^TM_C}$ explicitly since they differ 
significantly in appearance, particularly in the off diagonal elements 
which depend on the R-parity violating couplings:
\begin{equation}
\bf{M_CM_C^T}=\left[\matrix{
M^2+\half g^2v_2^2 & {1\over{\sqrt{2}}}g(Mv_1+\mu v_2) 
& {1\over\sqrt 2}g(Mv_3-\epsilon_3 v_2)\cr
{1\over{\sqrt{2}}}g(Mv_1+\mu v_2) 
& \mu^2+\half g^2v_1^2 +\half h_{\tau}^2v_3^2
& -\mu\epsilon_3+\half v_1v_3(g^2-h_{\tau}^2)  \cr
{1\over\sqrt 2}g(Mv_3-\epsilon_3 v_2) 
& -\mu\epsilon_3+\half v_1v_3(g^2-h_{\tau}^2)  
 & \half h_{\tau}^2v_1^2+\epsilon_3^2+\half g^2v_3^2 }\right]
\label{eq:MMt}
\end{equation}
and
\begin{equation}
\bf{M_C^TM_C}=\left[\matrix{
M^2+\half g^2(v_1^2+v_3^2) 
& {1\over{\sqrt{2}}}g(Mv_2+\mu v_1-\epsilon_3 v_3) & 0\cr
{1\over{\sqrt{2}}}g(Mv_2+\mu v_1-\epsilon_3 v_3) 
& \mu^2+\half g^2v_2^2+\epsilon_3^2 
& -{1\over \sqrt 2}h_{\tau}(\mu v_3+\epsilon_3v_1)   \cr
0 & -{1\over \sqrt 2}h_{\tau}(\mu v_3+\epsilon_3v_1)    
&  {1\over 2}h_{\tau}^2(v_1^2+v_3^2) }\right]
\label{eq:MtM}
\end{equation}
One notices a pair of zeros in the R--Parity violating elements of 
the matrix $\bf{M_C^TM_C}$. The second pair of elements which violate 
R--Parity turn out to be small. This can be seen from the fact that 
the term $(\mu v_3+v_1\epsilon_3)$ is naturally small since its square 
is proportional to the mass of $\nu_{\tau}$ 
\cite{chaHiggsEps,epsrad,epstalks}. Indeed, the above combination 
satisfy $\mu v_3+v_1\epsilon_3=\mu'v'_3$, where 
$\mu'=\sqrt{\mu^2+\epsilon_3^2}$ and $v'_3$ is the vacuum expectation
value (VEV) of the sneutrino field in a rotated basis where the 
$\epsilon_3$ term disappear from the superpotential (although 
reintroduced in the soft terms). In models with universality of soft
masses at the unification scale, this VEV $v'_3$ is radiatively 
generated and proportional to the bottom quark Yukawa coupling squared. 
On the other hand, the mixing between the neutralinos and the 
tau--neutrino in the rotated basis is proportional to $v'_3$, therefore 
a neutrino mass is 
induced and satisfy $m_{\nu_{\tau}}\sim v'^2_3$. In summary, 
$m_{\nu_{\tau}}\sim v'^2_3\sim (\mu v_3+v_1\epsilon_3)^2$ is naturally
small since it is one--loop induced and proportional to $h_b^2$. This
makes the R--Parity violating couplings $V_{31}$ and $V_{32}$ also
naturally small and controlled by $m_{\nu_{\tau}}$.

In contrast, the R-Parity violating elements in the matrix $\bf{M_CM_C^T}$ 
may be of greater magnitude. In this way, $U_{31}$ and $U_{32}$ may 
be larger than their similars $V_{31}$ and $V_{32}$. Nevertheless, a 
closer look tells us that in the limit 
$(\mu v_3+v_1\epsilon_3)\rightarrow 0$ (\ie, massless neutrino)
we find $U_{31}\rightarrow 0$. Therefore, $U_{31}$ is also small and
controlled by $m_{\nu_{\tau}}$. On the contrary, $U_{32}$ can be larger.
In the next section we show that this is not in conflict with the 
$\tau$ or $\nu_{\tau}$ couplings to gauge bosons.

\section{The $\tau$ couplings to gauge bosons}

The pair production of $\tau$ leptons in the SM proceeds via two tree
level Feynman diagrams, i.e., the s-channel mediated by a photon and a Z 
boson. In BRpV there is an extra diagram, namely that of t-channel
production mediated by a tau--sneutrino. This diagram arises because 
in general the $\tau$ has a gaugino component. 
One may attempt to constrain the chargino/tau mixing in 
eq.~(\ref{eq:ChaM6x6}) by using precision measurements at LEP1 of the 
$Z\tau\tau$ and $W\tau\nu_{\tau}$ couplings. The current experimental 
values of the axial vector ($g^{\tau}_A$) and the vector ($g^{\tau}_V$) 
parts of the $Z\tau\tau$ coupling are \cite{pdb}: 
\begin{equation}
g^{\tau}_A=-0.5009\pm0.0013,\qquad g^{\tau}_V=0.0374\pm0.0022
\label{bounds}
\end{equation}
whose tree level values are $g^{\tau}_A=-\half$ and 
$g^{\tau}_V=\half-2s_W^2$. It is costumary to write the coupling 
$Z\tau^+\tau^-$ in the MSSM in terms of the constants 
$g^{\tau}_L=-\half(g^{\tau}_A+g^{\tau}_V)$ and 
$g^{\tau}_R=\half(g^{\tau}_A-g^{\tau}_V)$ which are respectively the 
coupling strengths of the left and right handed $\tau$ to $Z$. 

In BRpV $g^{\tau}_L$ and $g^{\tau}_R$ are the third diagonal elements of 
the matrices $O'^L_{ij}$ and $O'^R_{ij}$, which are $3\times 3$ 
generalizations of the analogous $2\times 2$ matrices in the MSSM. 
If we call $F^{\pm}_i$, $i=1,2,3$, the three charged fermion mass
eigenstates, of which the first two are the charginos and the third one 
is the tau lepton, then the $ZF^+_jF^-_i$ interactions are
\begin{center}
\vspace{-50pt} \hfill \\
\begin{picture}(120,90)(0,25) 
\Photon(10,25)(60,25){4}{8}
\Vertex(60,25){3}
\ArrowLine(110,-5)(60,25)
\ArrowLine(60,25)(110,55)
\Text(85,-2)[]{$F^-_i$}
\Text(85,55)[]{$F^+_j$}
\Text(35,38)[]{$Z_{\mu}$}
\end{picture}
$
=i{g\over{2c_W}}\gamma^{\mu}\left[O'^L_{ij}(1-\gamma_5)
+O'^R_{ij}(1+\gamma_5)\right]
$
\vspace{30pt} \hfill \\
\end{center}
\vspace{10pt}
with the couplings $O'^L_{ij}$ and $O'^R_{ij}$ given by
\begin{eqnarray}
O'^L_{ij}&=&-V_{i1}V_{j1}^*-\half V_{i2}V_{j2}^*+\delta_{ij}s_W^2\,,
\nonumber\\
O'^R_{ij}&=&-U_{i1}^*U_{j1}-\half U_{i2}^*U_{j2}-\half U_{i3}^*U_{j3}
+\delta_{ij}s_W^2\,.
\label{OLRij}
\end{eqnarray}
In our notation the $\tau$ lepton is the ``third chargino'' and so to 
obtain the $Z\tau\tau$ coupling it is sufficient to put $i=j=3$. 
The axial part $g^{\tau}_A$ of the coupling $Z\tau\tau$ is given by 
$O'^R_{33}-O'^L_{33}$ while the vector part $g^{\tau}_V$ is given by 
$-O'^R_{33}-O'^L_{33}$. Written explicitly one finds:
\begin{eqnarray}  
g^{\tau}_A=-|U_{31}|^2-\half|U_{32}|^2-\half|U_{33}|^2
      +|V_{31}|^2+\half|V_{32}|^2\,,
\nonumber\\
g^{\tau}_V=|U_{31}|^2+\half|U_{32}|^2+\half|U_{33}|^2
      +|V_{31}|^2+\half|V_{32}|^2-2s_W^2\,.
\label{gvgaEps}
\end{eqnarray}
Of course, in the R-parity conserving limit, that is 
$V_{31}=V_{32}=U_{31}=U_{32}=0$, one recovers the formula for $\tau$ 
couplings in the MSSM. 

We have evaluated the numerical value of these couplings for $10^4$ 
randomly chosen points. All the points satisfy the following 
experimental mass limits: 
\begin{equation}
m_{\chi^+}\ge 70 \,{\mathrm{GeV}} \cite{tauJ},\quad m_{\nu_{\tau}}\le 18 
\,{\mathrm{MeV}} \cite{aleph97} \quad m_{\chi^0}\ge 20 \,{\mathrm{GeV}}\,.
\label{limits}
\end{equation} 
Note that, to be conservative, we applied assumed a lower bound on the
neutralino mass of 20 GeV. This is reasonable in view of the work
presented in ref. \cite{sensi}
\footnote{Strictly speaking, however, there is not yet a published 
determination on the neutralino bounds from LEP2 in the bilinear
model of broken Rp.}.  The couplings $g^{\tau}_A$ and $g^{\tau}_V$ are
functions of 6 independent parameters which are varied in the
following ranges:
\begin{eqnarray}
-200 \,{\mathrm{GeV}} \le& \mu, \epsilon_3 &\le 200 \,{\mathrm{GeV}}
\nonumber\\
-100 \,{\mathrm{GeV}} \le& v_3 &\le 100 \,{\mathrm{GeV}}
\nonumber\\
0.5 \le& \tan\beta &\le 90
\label{ranges}\\
30 \,{\mathrm{GeV}} \le& M &\le 200 \,{\mathrm{GeV}}
\nonumber\\
60 \,{\mathrm{GeV}} \le& m_{\tilde\nu} &\le 200 \,{\mathrm{GeV}}
\nonumber
\end{eqnarray}
\begin{figure}
\centerline{\protect\hbox{\psfig{file=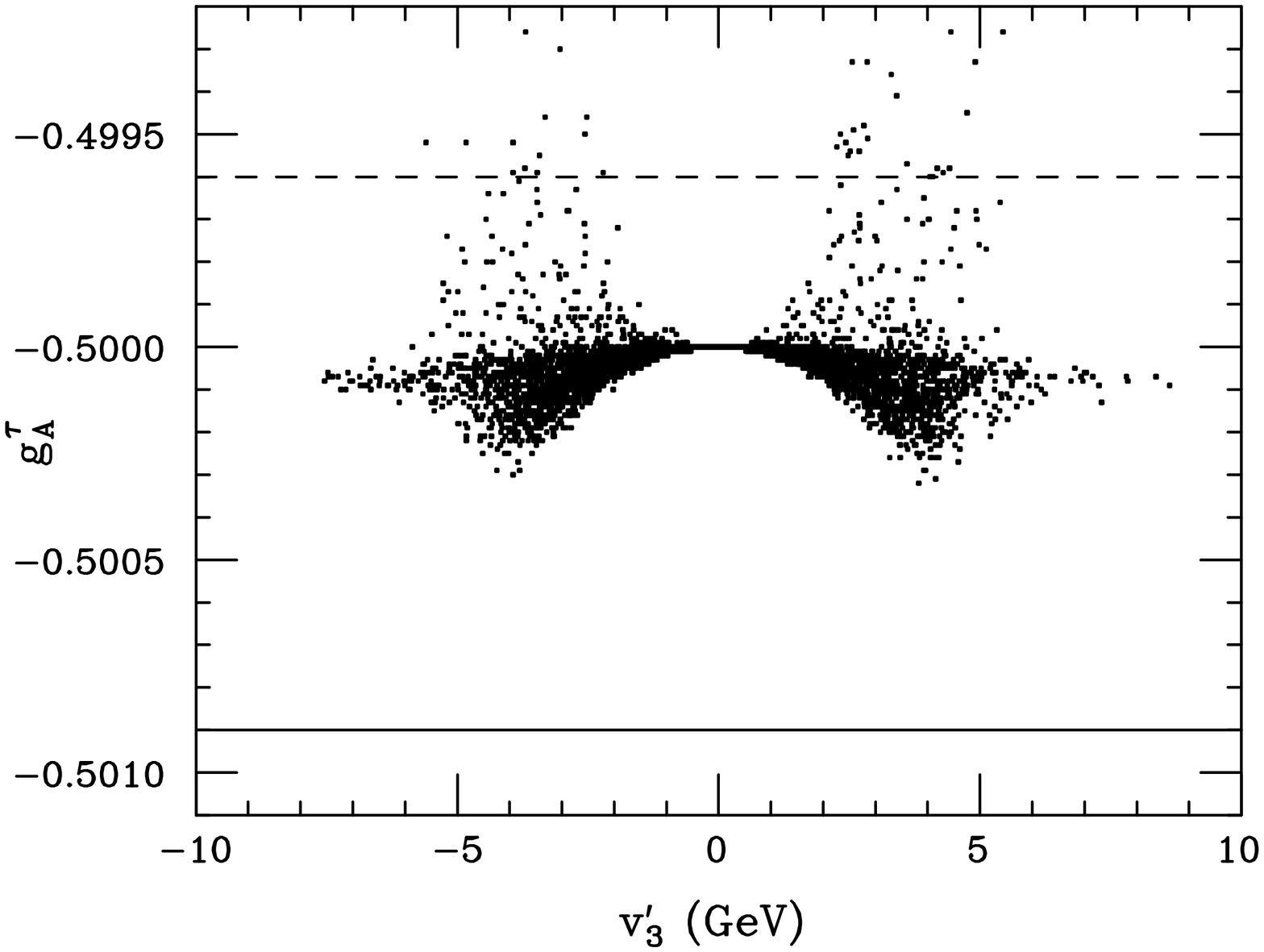,height=9cm,angle=90,width=0.80\textwidth}}}
\caption{Axial vector coupling of the $\tau$ to a $Z$ gauge boson as a 
function of the BRpV parameter $v'_3$. The solid line is the experimental
central value and the dashed line corresponds to $1\sigma$ deviation.}
\label{a_v3p}
\end{figure}
\begin{figure}
\centerline{\protect\hbox{\psfig{file=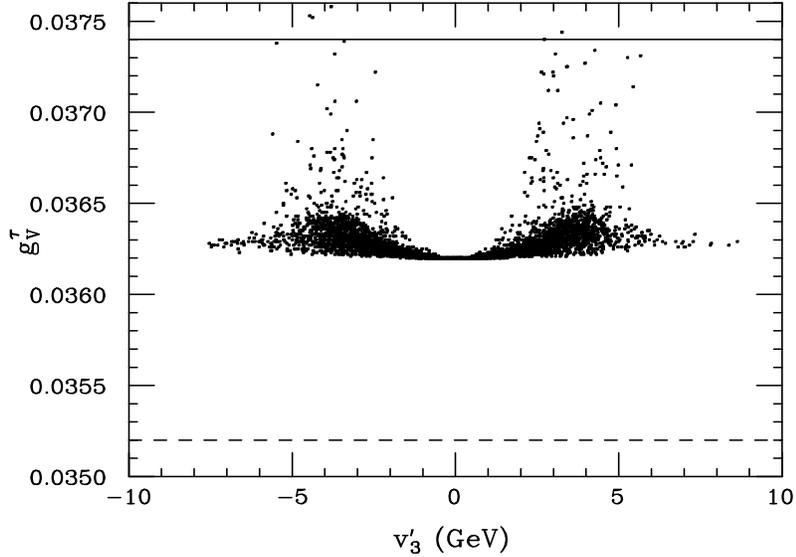,height=9cm,angle=90,width=0.80\textwidth}}}
\caption{Vector coupling of the $\tau$ to a $Z$ gauge boson as a 
function of the BRpV parameter $v'_3$. The solid line is the experimental
central value and the dashed line corresponds to $1\sigma$ deviation.}
\label{v_v3p}
\end{figure} 
In Fig.~\ref{a_v3p} we plot the axial vector coupling $g^{\tau}_A$ as
a function of the sneutrino vacuum expectation value $v'_3$ in the
rotated basis. The central value is given by $g^{\tau}_A=-0.5009$
(solid line) and the horizontal dashed line corresponds to $1\sigma$
deviation. Clearly, the great majority of the points fall within the
LEP1 bound on $g^{\tau}_A$ at the $1\sigma$ level. Note that the values
of $\epsilon_3$ and $v_3$ can be large and constrained only by
$v'_3\lsim 10$ GeV. Similarly, in Fig.~\ref{v_v3p} we plot the vector
coupling $g^{\tau}_V$ as a function of $v'_3$. The solid horizontal
line at $g^{\tau}_V=0.0374$ is the central value, and the dashed line
denotes the $1\sigma$ deviation.  In this case all points fall inside
the $1\sigma$ region.

Let us now consider how the tau mixing with the charginos
affects the $W\tau\nu_{\tau}$ vertex. Any change in this coupling would
affect the $\tau$ decay widths which are measured very accurately.
In our model there are five neutral fermions mass eigenstates which we
denote $F^0_1$, $i=1,...5$, with the lightest (corresponding to $i=5$) 
being the tau neutrino and the first four being the ``neutralinos''.
The $5\times 5$ neutralino mass matrix is diagonalized by the matrix 
${\bf N}_{ij}$. The vertex $WF^{\pm}_jF^0_i$ is 
\begin{center}
\vspace{-50pt} \hfill \\
\begin{picture}(120,90)(0,25) 
\Photon(10,25)(60,25){4}{8}
\Vertex(60,25){3}
\ArrowLine(110,-5)(60,25)
\ArrowLine(60,25)(110,55)
\Text(85,-2)[]{$F^0_i$}
\Text(85,55)[]{$F^+_j$}
\Text(35,38)[]{$W_{\mu}$}
\end{picture}
$
=i{g\over 2}\gamma^{\mu}\left[O^L_{ij}(1-\gamma_5)
+O^R_{ij}(1+\gamma_5)\right]
$
\vspace{30pt} \hfill \\
\end{center}
\vspace{10pt}
with the generalized $O^L$ and $O^R$ couplings given by
\begin{eqnarray}
O^L_{ij}&=&-{\textstyle{1\over \sqrt 2}}N_{i4}V_{j2}^*+N_{i2}V_{j1}^*\,,
\nonumber\\
O^R_{ij}&=&{\textstyle{1\over \sqrt 2}}N_{i3}^*U_{j2}+N_{i2}^*U_{j1}+
{\textstyle{1\over \sqrt 2}}N_{i5}^*U_{j3}\,.
\label{OlOr}
\end{eqnarray}
For the vertex $W\tau\nu_{\tau}$ one sets $i=5$ and $j=3$. In the 
Rp-conserving limit one would obtain 
\begin{eqnarray}
-{\textstyle{1\over \sqrt 2}}N_{54}V_{32}^*+N_{52}V_{31}^*
&\longrightarrow& 0\,,
\nonumber\\
{\textstyle{1\over \sqrt 2}}N_{53}^*U_{32}+N_{52}^*U_{31}+
{\textstyle{1\over \sqrt 2}}N_{55}^*U_{33}
&\longrightarrow& {\textstyle{1\over \sqrt 2}}\,.
\label{Wtauneu}
\end{eqnarray}
With the same $10^4$ randomly chosen points we find that the deviations 
from the SM values of the $W\tau\nu_{\tau}$ couplings are well inside the 
experimental error, and less pronounced than the corresponding deviations
for the $Z\tau\tau$ vertex.

The reason why the deviations of the tau couplings to gauge bosons
are small with respect to the SM predictions can be traced to two facts. 
First the tau--neutrino mass is small, and second the Higgs superfield
$H_1$ and the tau--lepton superfield $L_3$ both possess the same SU(2) 
quantum numbers. 

Summarizing the results of this section, we conclude that the
couplings $Z\tau\tau$ and $W\tau\nu_{\tau}$ can be easily maintained
within the experimental bounds even for large values of $\epsilon_3$
and $v_3$.  This has important consequences for the phenomenology of
the BRpV model in general \cite{beyonddesert}, with the most salient
effect being the existence of Rp violating branching ratios for
sparticles, the most dramatic example being the LSP. Moreover, Rp
violation may also affect the expected mass spectrum in comparision
with the Rp conserving case. One example recently considered in
ref.~\cite{chaHiggsEps} is the charged Higgs mass can be lowered below
80 GeV if $\epsilon_3$ is large.

\section{Single Chargino Production}

In this section we consider the single chargino production in 
electron--positron annihilation. We study in turn the phenomenology
at LEP1, LEP2, and NLC ($\sqrt{s}=500$ GeV). In general the total 
cross--section consists of three distinct contributions given by:
\begin{equation}
\sigma(e^+e^-\rightarrow Z, \tilde\nu_{\tau}\rightarrow
\tilde\chi^{\pm}_1\tau^{\mp})=\sigma_{Z}+\sigma_{\tilde\nu}
+\sigma_{\tilde\nu Z}
\label{crossSix}
\end{equation}
Note that an intermediate photon does not contribute. Explicit 
expressions for these formulae can be found in ref.~\cite{Bartletal}. 
The generalization to BRpV is straightforward and is obtained by 
replacing the $2\times 2$ matrices $O'^L$, $O'^R$, $V$ and $U$ by 
$3\times 3$ matrices and summing over three ``charginos''.

First, we consider the single chargino production at LEP1. In this
case, the terms involving $\tilde\nu_{\tau}$ are negligible, which is
expected at the $Z$ peak. In addition, the sneutrino contribution is
proportional to $|V_{31}|^2$ which is small, as we mentioned in the
previous section.

The magnitude of $\sigma_{Z}$ depends crucially on the parameters 
$O'^L_{13}$ and $O'^R_{13}$ which are given by
\begin{eqnarray} 
O'^L_{13}&=&-V_{11}V_{31}^*-\half V_{12}V_{32}^*
\nonumber\\
O'^R_{13}&=&-U_{11}^*U_{31}-\half U_{12}^*U_{32}-\half U_{13}^*U_{33}
\label{Op13}
\end{eqnarray}
Since $O'^L_{13}$ depends on the R--parity violating $V_{31}$ and $V_{32}$ 
elements (which are small) one would expect the potentially larger 
contribution to come from $O'^R_{13}$, since the element $U_{32}$ may 
have larger values. Nevertheless, it is possible to show that in the
limit $m_{\nu_{\tau}}\rightarrow 0$ we get 
$U_{12}^*U_{32}-U_{13}^*U_{33}\rightarrow 0$. Therefore, the whole
cross section is controlled by the neutrino mass. Numerically, we find 
that the elements $V_{31}$ and $V_{32}$ have a greater contribution to 
the cross--section.

At LEP1 despite the small values for the couplings $O'^L_{13}$ and
$O'^R_{13}$ and the relatively low luminosity of 82 pb$^{-1}$, we
benefit from being at the $Z$ peak. A study of the cross-section at
LEP1 was carried out in Ref.~\cite{ConchaV,chitau} and it was found
maximum cross-sections of order 1 pb, using lighter values for the
chargino mass (45 GeV) and allowing a heavier tau neutrino (35
MeV). We improve that discussion by including a full scan of the
parameter space as well as imposing the latest limits on the sparticle
masses as well as $m_{\nu_{\tau}}$. The latest mass limits [see
eq.~(\ref{limits})] restrict the parameter space for large
cross-sections, so much so that the results of
Ref.~\cite{chitau,ConchaV} suggest that only cross--sections of order
0.1 pb are now possible, which would be unobservable at LEP1. Here we
show that this was underestimated and explain the reasons.

\begin{figure}
\centerline{\protect\hbox{\psfig{file=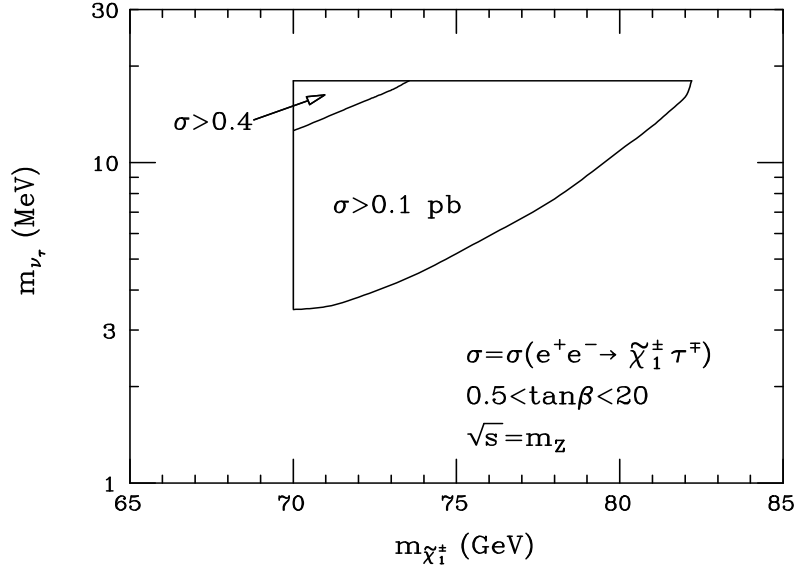,height=9cm,angle=90,width=0.80\textwidth}}}
\caption{Regions of attainable cross section in BRpV in the plane 
tau neutrino mass v/s chargino mass for moderate values of $\tan\beta$.}
\label{ntch_20}
\end{figure}
Fig.~\ref{ntch_20} shows regions of attainable cross--sections in the 
$m_{{\nu}_\tau}-m_{\chi_1}$ plane. We consider $\tan\beta\le 20$ and 
so is an update of the figure 7 in Ref.~\cite{ConchaV}. The reason why
we get larger cross sections in this figure with respect to the previous 
reference is that we consider larger values of $\epsilon_3$. Inside 
the largest triangular region lie the points with $\sigma>0.1$ pb, and 
inside the smaller triangular region we have $\sigma>0.4$ pb, values 
which are in the limit of observability at LEP1. The vertical and 
horizontal sides of the triangles correspond to two of the limits in 
eq.~(\ref{limits}).

\begin{figure}
\centerline{\protect\hbox{\psfig{file=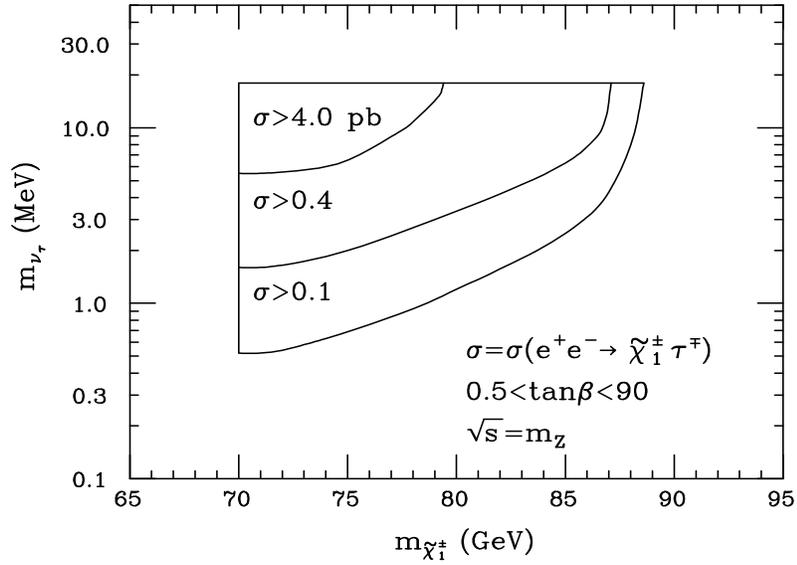,height=9cm,angle=90,width=0.80\textwidth}}}
\caption{Regions of attainable cross section in BRpV in the plane 
tau neutrino mass v/s chargino mass including large values of $\tan\beta$.}
\label{ntch_90}
\end{figure}
In Fig.~\ref{ntch_90} we consider $\tan\beta$ up to $90$ and this allows
much larger cross--sections, for a given tau neutrino mass. The reason
is that the cross section depends crucially on the tau Yukawa
coupling: larger values of $\tan\beta$ increase the value of
$h_{\tau}$ and this in turn increases $\sigma$. This region of large
$\tan\beta$ was not explored in ref.~\cite{ConchaV}. In this figure we
have three regions with the total cross section larger than 0.1, 0.4,
and 4 pb.

\begin{figure}
\centerline{\protect\hbox{\psfig{file=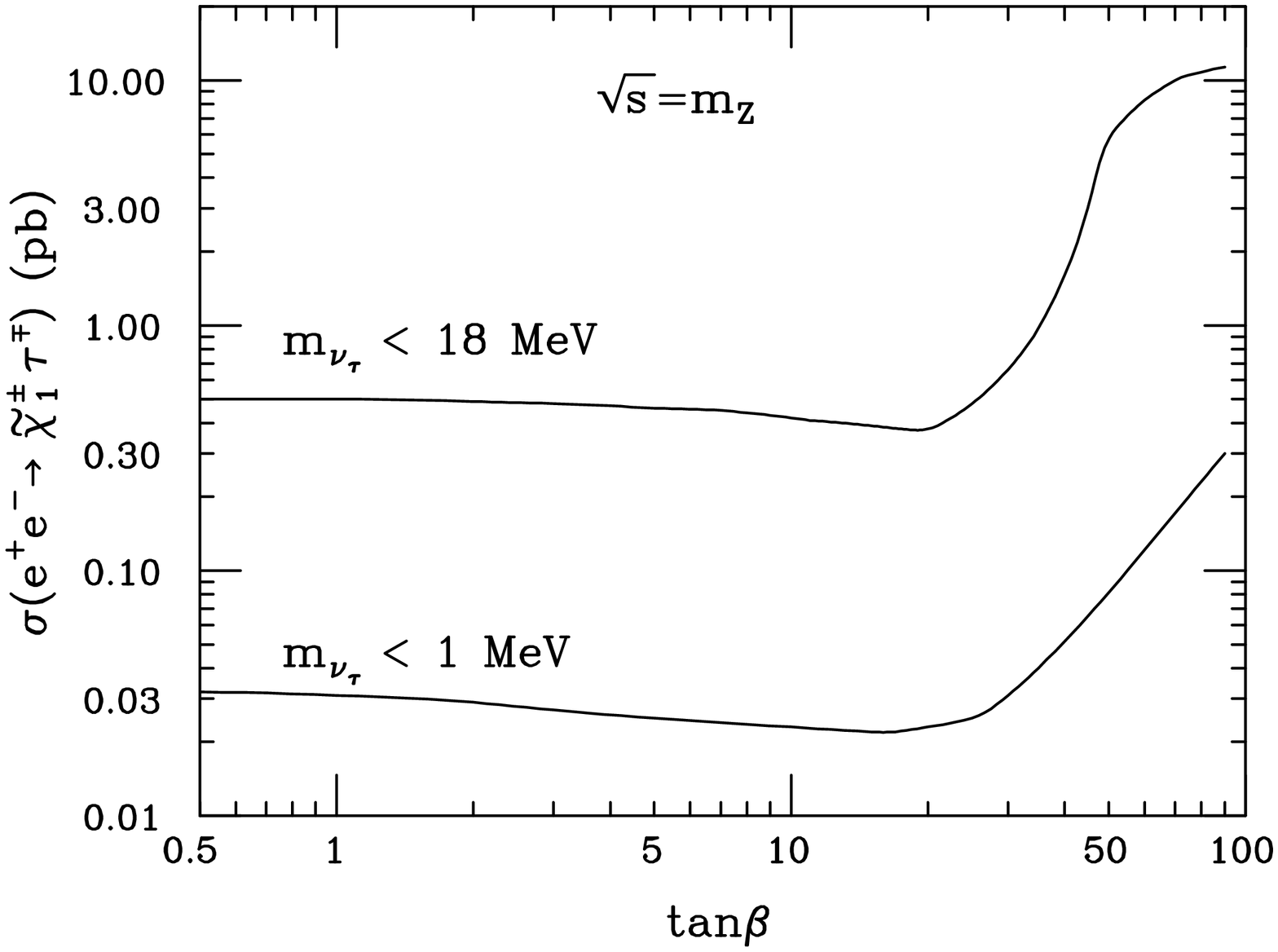,height=9cm,angle=90,width=0.80\textwidth}}}
\caption{Maximum single chargino production cross section in BRpV as a 
function of $\tan\beta$ for two different upper bounds on the tau 
neutrino mass.}
\label{cstb}
\end{figure}
To understand better the relation between the cross section and $\tan\beta$
we show in Fig.~\ref{cstb} the explicit cross--section dependence on 
$\tan\beta$. It shows clearly a steep climb for $\tan\beta\gsim 30$.
In addition we show two curves, one for $m_{\nu_{\tau}}<18$ MeV and
another one with $m_{\nu_{\tau}}<1$ MeV. They clearly show that the
cross section is controlled by $m_{\nu_{\tau}}$ and will approach
zero as the neutrino mass goes to zero, as expected in the model,
see, e.g. discussions given in \cite{beyonddesert}.

\begin{figure}
\centerline{\protect\hbox{\psfig{file=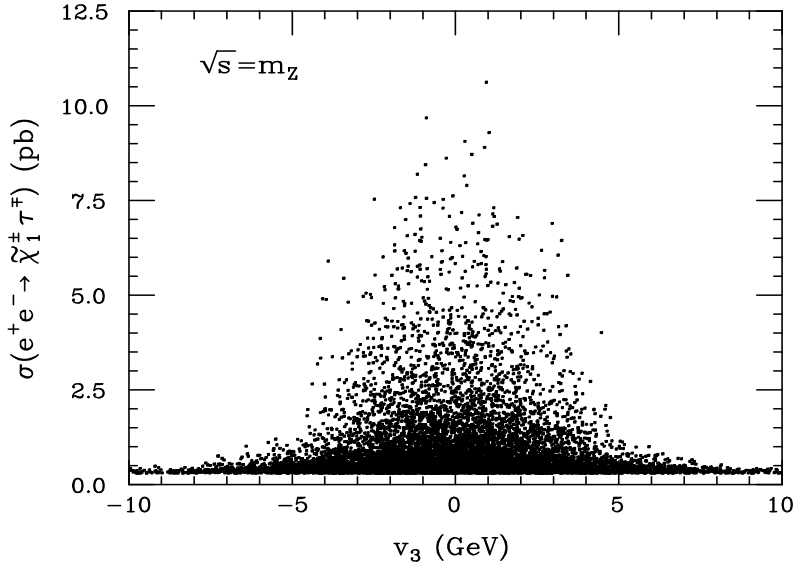,height=9cm,angle=90,width=0.80\textwidth}}}
\caption{Maximum single chargino production cross section in BRpV as a 
function of $v_3$.}
\label{sigmav3}
\end{figure}

Cross-sections as large 10 pb can be obtained for $\tan\beta\to 90$.
Of interest is the region $55\lsim\tan\beta\lsim60$, which is required
if top--bottom--tau Yukawa coupling unification is imposed
\cite{yukeps}. Within this region we find maximum single chargino 
production cross section values of 8 pb. The cross-section has a
direct dependence on the Tau Yukawa coupling and this can be explained
as follows. As mentioned above, the coupling $O'^L_{13}$ gives the
major contribution to the cross--section and $O'^L_{13}$ is a function
of $V_{31}$ and $V_{32}$, which in turn have a strong dependence on
$h_{\tau}$. One can infer from the matrix $M_C^TM_C$ that the largest
values of $h_{\tau}$ require low $v_3$ and large $\tan\beta$, i.e., if
the term $v_1^2+v_3^2$ is made smaller then $h_{\tau}$ must be
increased to keep the $\tau$ mass at $1.777$ GeV.  Since $h_{\tau}$
takes its largest values for $v_3=0$ and large $\tan\beta$ we can
explain the $\tan\beta$ dependence found in the cross--section in
Fig.~\ref{cstb}. We note that in Ref.~\cite{ConchaV} $v_3$ was equated
to zero, although here we explicit show that $v_3=0$ maximizes the
cross--section. This can seen explicitly in a graphic form as depicted
in Fig.~\ref{sigmav3}. 
This result has important implications if one considers the BRpV model
as a low energy form of a spontaneous Rp violating model, in which a
massless majoron is present. Such models require $v_3\le 100$ MeV. Our
past discussion as well as Fig.~\ref{sigmav3} shows that it is
precisely for such small $v_3$ values that we have maximum
cross--sections. In addition the presence of a majoron would allow one
to use the laboratory bound on $m_{\tau_{\nu}}$ which is less
stringent than the nucleosynthesis bound, thus allowing
cross--sections of 1 to 10 pb in a limited range of parameter
space. In the case of no majoron, one is obliged to use the tighter
bound $m_{{\nu}_\tau} \lsim 2$ MeV, which only allows maximum
cross--sections of order 0.4 pb at LEP1.

At LEP2 one moves away from the $Z$ peak and so the cross-section falls.
With $4\times 10^4$ random points chosen the maximum cross-section that 
we found had a value of 7.4 fb, and so 3.7 events would be expected at 
LEP2 with a luminosity of 500 ${\rm pb}^{-1}$. Hence we conclude that 
LEP2 has no chance of obtaining a signal in this channel. 

\begin{figure}
\centerline{\protect\hbox{\psfig{file=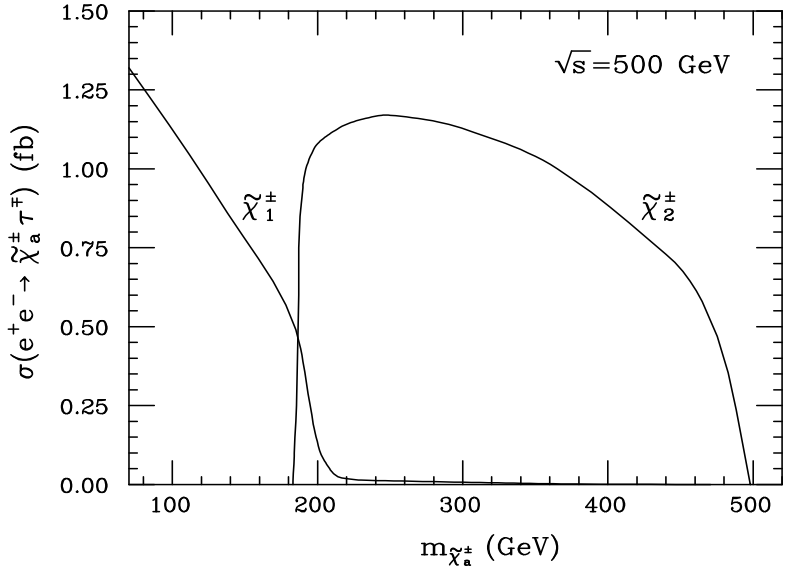,height=9cm,angle=90,width=0.80\textwidth}}}
\caption{Maximum single chargino production cross section as a function 
of the chargino mass at NLC in BRpV. Light and heavy charginos are 
displayed.}
\label{csmc}
\end{figure}
At an NLC of $\sqrt{s}=500$ GeV one finds even smaller cross--sections
although we benefit from the higher luminosity ($30\to 100$
fb${}^{-1}$).  Production of the heavier chargino $\chi_2$ is now
possible and we show in Fig.~\ref{csmc} the maximum values of the
cross--section of both $\tilde\chi_1$ and $\tilde\chi_2$ as a function
of chargino mass. We find that for $\tilde\chi_1$ the cross section
starts at $m_{\chi_1}=70$ GeV with 1.3 fb and then drops. In addition,
the cross section for $\tilde\chi_2$ peaks at around $m_{\chi_2}=250$
GeV with 1.2 fb and then falls. One would expect a smaller cross
section for the heavier chargino since its mass is larger, but this is
not the case. One can explain this as follows. For the cross-section
of $\tilde\chi_1$ there exists a cancellation between the terms
$\sigma_{\tilde\nu}$ and $\sigma_{Z\tilde\nu}$, which are of similar
magnitude (each one is larger than $\sigma_Z$ and with opposite
relative sign). Hence one is left with the contribution $\sigma_Z$
which gives the shape in Fig.~\ref{csmc}. For $\tilde\chi_2$ the story
is different. We find numerically that the coupling $O'^L_{23}$ is
smaller than $O'^L_{13}$. This implies that $\sigma_{Z}$ and
$\sigma_{Z\tilde\nu}$ are smaller in the case of $\tilde\chi_2$
compared with $\tilde\chi_1$.  Therefore the shape of the
cross-section in Fig.~\ref{csmc} for $\tilde\chi_2$ is dictated by the
t-channel contribution $\sigma_{\tilde\nu}$.

\section{Discussion and Conclusions}

We have studied the charged and neutral current couplings of the tau
lepton in models with spontaneous or bilinear breaking of
R--parity. We showed that precision measurements of the $Z\tau\tau$
coupling at LEP1 allow relatively large values of the effective Rp
breaking parameter $\epsilon_3$ ($\to 200$ GeV) since such values only
induce large mixing between $\tau^-$ and $\tilde H_1^-$ which share
the same $SU(2)$ quantum numbers.  We then moved to see the possible
manifestations of Rp breaking as single production of charginos in
models with spontaneous or bilinear breaking of R--parity at various
$e^+ e^-$ accelerators such as LEP1, LEP2, and a NLC with
$\sqrt{s}=500$ GeV. The numerical values of the cross-sections are
small at LEP2 and at NLC, with maximum values of around 7.4 fb and 1.3
fb respectively. With large luminosity at the latter (say, 30
fb${}^{-1}$ or greater) there would be some chance of detection in
this channel. On the contrary, the signal is un--observable at LEP2.
At LEP1 prospects are vastly improved, and we found that the large
values $\tan\beta$ ($\approx 90$) allow cross-sections as great as 10
pb. Assuming top-bottom-tau Yukawa unification (which requires
$55\lsim\tan\beta\lsim 60$) we find maximum cross-sections of around
$8$ pb.

The cross section decreases with decreasing $m_{\nu_{\tau}}$. In
addition, there is a direct correlation between the tau Yukawa
coupling ($h_{\tau}$) and the cross-section, with large values of the
latter only possible for large values of the former. The coupling
$h_{\tau}$ itself has a strong dependence on $v_3$ and $\tan\beta$,
with its greatest values being found at $v_3=0$ and large $\tan\beta$.

Our results are to a large extent model independent, since they depend
only of the structure of the chargino-tau mass matrix and this is 
universal in models with spontaneous breaking of R--parity as well as 
their effective truncation in terms of a bilinear explicit Rp violating
superpotential term (BRpV model). The model dependence arises due to
the following:
\begin{enumerate}
\item
The bounds on $m_{\nu_\tau}$. This depends on whether or not the
majoron exists. As already mentioned, if it does, we should apply the
laboratory limit from LEP1, which allows for example, much larger
single chargino production cross sections at the Z peak, due to the
strict correlation between Rp violation observables and
$m_{\nu_\tau}$.  In models without the majoron, such as those in ref.
\cite{ConchaV,finnish}, or the BRpV model \cite{e3others,epsrad} one
must apply the tighter limits of about one MeV to the tau neutrino mass
that follows from primordial Big-Bang nucleosynthesis and we loose the
large cross sections.
\item
The bounds on $m_{\tilde\chi_1}$. This also depends on whether or not
the majoron exists. If it does, we should note that the chargino may
be lighter than in the MSSM, the lower bound being about 70 GeV from
LEP2 data \cite{tauJ}, and therefore the left part in figures 3 and
4 is justified. If the majoron is absent, then we expect a tighter
limit on the lighter chargino mass, similar to that which holds in
the MSSM, though, strictly speaking, there has been no dedicated
chargino search taking into account the decay modes expected in models
with broken R--parity which would affect at least its cascade decays,
since the lightest neutralino would decay.
\item
The allowed values of $v_3$. This also depends on whether or not the
majoron exists. If it does, we should note that in order to avoid
exceesive stellar cooling by majoron emmission one needs small
$v_3$. Fortunately, as seen in Fig.~\ref{sigmav3} our results for
single production at LEP1 are better for small $v_3$.  
\end{enumerate}
We conclude that the most favourable model which allows a sizeable
single chargino cross section at the Z peak is the simplest
$SU(2)\times U(1)$ spontaneously broken R--parity model, which is
characterized by the existence of a majoron and a strict correlation
between R--parity breaking observables and neutrino mass. Possible
signatures will be model-dependent. For example, in the majoron
version there will be di-tau + missing momentum, coming from the
$e^+e^-\to\chi\tau$ with $\chi \to \tau + J$, which has been discussed
in ref. \cite{tauJ} in the context of chargino pair-production at
LEP2. In the explicit BRpV version the main effect of Rp violation is
in the lightest neutralino decay which would lead to a rich cascade
decay pattern characterized by large fermion multiplicities.

\vskip 1cm

This work was supported by DGICYT under grants PB95-1077 and by the
TMR network grant ERBFMRXCT960090 of the European Union. M. A. D. was
supported by a DGICYT postdoctoral grant and A. Akeroyd by a CSIC-UK
Royal Society fellowship.

\end{document}